\documentclass[12pt]{article}
\usepackage{amsmath}
\usepackage{amsfonts}
\usepackage{graphicx}
\usepackage{amssymb}
\usepackage[unicode,bookmarks,bookmarksopen,bookmarksopenlevel=2,colorlinks,linkcolor=blue,citecolor=green]{hyperref}

\def\beq{\begin{equation}}
\def\eeq{\end{equation}}
\def\bearr{\begin{eqnarray}}
\def\eearr{\end{eqnarray}}

\begin{document}

\title{\textbf{Generalised hydrodynamic reductions of the kinetic equation
for soliton gas}}
\author{Gennady A.~El$^{1}$, Maxim V. Pavlov$^{2,3}$, Vladimir B. Taranov$%
^{4}$  \and   \\
$^{1}$Department of Mathematical Sciences,\\
Loughborough University, Loughborough, Leicestershire, UK\\
\\
$^{2}$Department of Mathematical Physics,\\
P.N. Lebedev Physical Institute of Russian Academy of Sciences,\\
Moscow, 53 Leninskij Prospekt, Moscow, Russia\\
\\
$^{3}$Laboratory of Geometric Methods in Mathematical Physics,\\
Department of Mech. \& Math., Moscow State University,\\
1 Leninskie gory, Moscow, Russia\\
\\
$^{4}$Institute for Nuclear Research,\\
National Academy of Sciences of Ukraine,\\
47 pr. Nauky, Kyiv, Ukraine}
\date{}
\maketitle

\begin{abstract}
We derive generalised multi-flow hydrodynamic reductions of the nonlocal
kinetic equation for a soliton gas and investigate their structure. These
reductions not only provide further insight into the properties of the new
kinetic equation but also could prove to be representatives of a novel class
of integrable systems of hydrodynamic type, beyond the conventional
semi-Hamiltonian framework.
\end{abstract}


\section{Introduction}

The generalised soliton-gas kinetic equation represents an
integro-differential system \cite{kinetic}
\begin{equation}
f_{t}+(sf)_{x}=0,  \label{a}
\end{equation}%
\begin{equation}
s(\eta )=S(\eta )+\frac{1}{\eta }\int\limits_{0}^{\infty }G(\eta ,\mu )f(\mu
)[s(\mu )-s(\eta )]d\mu . \,  \label{kin1}
\end{equation}
Here $f(\eta )\equiv f(\eta ,x,t)$ is the distribution function and $s(\eta
)\equiv s(\eta ,x,t)$ is the associated transport velocity. The (given)
functions $S(\eta )$ and $G(\eta ,\mu )$ do not depend on $x$ and $t$. The
function $G(\eta ,\mu )$ is assumed to be symmetric, i.e. $G(\eta ,\mu
)=G(\mu ,\eta )$.

System (\ref{a}), (\ref{kin1}) with
\begin{equation}
S(\eta )=4\eta ^{2}\,,\qquad G(\eta ,\mu )=\log \left| \frac{\eta -\mu }{%
\eta +\mu }\right|  \label{kdv}
\end{equation}
was derived in \cite{el03} as an infinite-genus thermodynamic limit of the
Whitham modulation equations associated with the KdV equation, $\varphi
_{t}-6\varphi \varphi _{x}+\varphi _{xxx}=0$, and was shown to describe
macroscopic dynamics of a soliton gas, a disordered infinite-soliton
ensemble of finite density \cite{ekmv99}. In the KdV context, $\eta \geq 0$
is a real-valued spectral parameter and the function $f(\eta ,x,t)$ is the
distribution function of solitons over the spectrum so that $\kappa
=\int_{0}^{\infty }f(\eta )d\eta =\mathcal{O}(1)$ is the spatial density of
solitons. If $\kappa \ll 1$, the first order approximation of (\ref{kin1}), (%
\ref{kdv}) yields Zakharov's kinetic equation for a dilute gas of KdV
solitons \cite{zakh71}. The quantity $S(\eta )$ in (\ref{kin1}) has a
natural meaning of the velocity of an isolated (free) soliton with the
spectral parameter $\eta $ and the function $G(\eta ,\mu )/\eta $ is the
expression for a phase shift of this soliton occurring after its collision
with another soliton having the spectral parameter $\mu <\eta $. Then $%
s(\eta ,x,t)$ acquires the meaning of the self-consistently defined mean
local velocity of solitons with the spectral parameter close to $\eta $. A
straightforward physical derivation of the kinetic equation (\ref{a}), (\ref%
{kin1}) for integrable systems, based on the original Zakharov \cite{zakh71}
phase-shift reasoning was proposed in \cite{elkam05}.

In recent paper \cite{kinetic}, the multi-flow hydrodynamic reductions of
the kinetic equation (\ref{a}), (\ref{kin1}) were studied using the
so-called `cold-gas' ansatz
\begin{equation}
f(\eta ,x,t)=\sum_{m=1}^{N}f^{m}(x,t)\delta (\eta -\eta ^{m}),  \label{cold}
\end{equation}%
where the `spectral' components $\eta ^{N}>\eta ^{N-1}>\dots >\eta ^{1}>0$
are arbitrary numbers. These `isospectral' cold-gas reductions were shown to
have the form of systems of hydrodynamic conservation laws%
\begin{equation}
u_{t}^{i}=(u^{i}v^{i})_{x}\,,\qquad i=1,\dots ,N\,,  \label{01}
\end{equation}%
where the conserved `densities' $u^{i}=\eta ^{i}f(\eta ^{i},x,t)$ and the
associated velocities $v^{i}=-s(\eta ^{i},x,t)$ are related algebraically:%
\begin{equation}
v^{i}=\xi _{i}+\sum_{m\neq i}\epsilon ^{im}u^{m}(v^{m}-v^{i})\,,\quad
\epsilon^{ik}=\epsilon^{ki}\,.  \label{02}
\end{equation}%
Here
\begin{equation}  \label{03}
\xi _{i}=-S(\eta ^{i})\,,\qquad \epsilon ^{ik}=\frac{G(\eta ^{i},\eta ^{k})}{%
\eta ^{i}\eta ^{k}}\,,\qquad i\neq k\,.
\end{equation}%
The isospectral cold-gas reductions (\ref{01}), (\ref{02}) were proven in
\cite{kinetic} to represent integrable (semi-Hamiltonian \cite{ts91})
linearly degenerate hydrodynamic type systems (see \cite{fer91}, \cite{pav87}%
) for arbitrary $N$, which is a strong indication that the full kinetic
equation (\ref{a}), (\ref{kin1}) could constitute an integrable system in
the sense yet to be explored.

The present paper is devoted to a more general multi-flow hydrodynamic
approximation of the kinetic equation (\ref{a}), (\ref{kin1}), which we
derive by considering an ansatz (see, for instance, \cite{Silin})
\begin{equation}
f(\eta ,x,t)=\sum_{m=1}^{N}f^{m}(x,t)\delta (\eta -\eta ^{m}(x,t))
\label{05}
\end{equation}
with the `spectral components' $\eta^k=\eta ^{k}(x,t)$ being (unknown)
functions of $x$ and $t$ rather than arbitrary constants as in (\ref{cold}).
We show that the corresponding $N$-flow non-isospectral hydrodynamic
reductions have the form of $2N$-component hydrodynamic type systems
\begin{equation}  \label{r0}
u_{t}^{i}=(u^{i}v^{i})_{x},\text{ \ \ }\eta _{t}^{i}=v^{i}\eta _{x}^{i},
\quad i=1,2, \dots, N,
\end{equation}%
where the functions $u^i(x,t)$, $v^i(x,t)$ and $\eta^i(x,t)$ are related
algebraically by the same equations (\ref{02}), (\ref{03}) provided certain
restrictions on the behaviour of the kernel function $G(\eta, \mu)$ for $%
\eta \to \mu$ are satisfied.

System (\ref{r0}), (\ref{01}), (\ref{02}) is not integrable by the standard
Tsarev generalized hodograph method, because it possesses just $N$ Riemann
invariants and has double characteristic velocities. However, having in mind
that this system is obtained as an exact reduction of an integrable system
(at least for $S(\eta)$, $G(\eta, \mu)$ defined by (\ref{kdv}) --- the KdV
case), one can expect that the multi-flow reductions (\ref{r0}) will be
integrable by some modification of the generalised hodograph method \cite%
{ts91}. This could lead to an extension of the conventional notion of an
integrable system of hydrodynamic type. We are going to investigate this
problem in detail in future publications.

\section{Generalised hydrodynamic reductions}

\subsection{Evolution equations}

Substituting (\ref{05}) into (\ref{a}) we obtain (hereafter we shall be
using a shorthand notation $\eta ^{i}$ for $\eta ^{i}(x,t)$)%
\begin{equation*}
\frac{\partial }{\partial t}\left( \sum\limits_{i=1}^{N}f^{i}(x,t)\delta
(\eta -\eta ^{i})\right) +\frac{\partial }{\partial x}\left( s(\eta
,x,t)\sum\limits_{i=1}^{N}f^{i}(x,t)\delta (\eta -\eta ^{i})\right) =0,
\end{equation*}%
Differentiating and collecting the terms for $\delta (\eta -\eta ^{i})$ and $%
\delta ^{\prime }(\eta -\eta ^{i})$ we obtain
\begin{equation}
\sum\limits_{n=1}^{N}[f_{t}^{n}+(s(\eta ,x,t)f^{n})_{x}]\delta (\eta -\eta
^{n})-\sum\limits_{n=1}^{N}[f^{n}\eta _{t}^{n}+s(\eta ,x,t)f^{n}\eta
_{x}^{n}]\delta ^{\prime }(\eta -\eta ^{n})=0.  \label{d}
\end{equation}%
Here $f^i \equiv f^i(x,t)$. Evaluating asymptotic behavior of this
expression near each point $\eta ^{i}$ we arrive $2N$ component hydrodynamic
type system (cf. (\ref{01}))%
\begin{equation}
f_{t}^{i}+(s(\eta ^{i},x,t)f^{i})_{x}=0,\text{ \ }\eta _{t}^{i}+s(\eta
^{i},x,t)\eta _{x}^{i}=0, \quad n=1, \dots, N\, .  \label{c}
\end{equation}%
It is instructive to derive the hydrodynamic reduction (\ref{c}) by a direct
calculation. This is done by integrating (\ref{d}) with respect to $\eta$
over a small vicinity of each point $\eta = \eta^{i}$ with the weights $1$
and $(\eta -\eta ^{i})$, respectively.

Let us fix $x=x_0$. Then, assuming the ordering $\eta ^{N}>\eta ^{N-1}>\dots
>\eta ^{1}>0$ to hold for all $t$ in a small vicinity of $x_0$ we introduce $%
N$ closed intervals $\sigma_i=[\eta^i-\varepsilon_i, \eta^i + \varepsilon_i]$
choosing $\varepsilon_i>0$ in such a way that in the vicinity of $x_0$ one
has $\eta^j (x,t) \in \sigma_i$ iff $j=i$.

We now integrate (\ref{d}) over $\sigma_i$:
\begin{equation*}
\int \limits_{\sigma_i} \left[ \sum\limits_{n=1}^{N}[f_{t}^{n}+(s(\eta
,x,t)f^{n})_{x}]\delta (\eta -\eta ^{n})-\sum\limits_{n=1}^{N}[f^{n}\eta
_{t}^{n}+s(\eta ,x,t)f^{n}\eta _{x}^{n}]\delta ^{\prime }(\eta -\eta ^{n})%
\right] d\eta =0,
\end{equation*}
which reduces, after integrating the term with $\delta^{\prime i})$ by
parts, to
\begin{equation}  \label{int1}
\int \limits_{\sigma_i}\left[ \sum\limits_{n=1}^{N} [f_{t}^{n} + s(\eta
,x,t)f_{x}^{n}+\frac{\partial s(\eta ,x,t)}{\partial x}f^{n}+\frac{\partial
s(\eta ,x,t)}{\partial \eta }f^{n}\eta _{x}^{n} ] \delta (\eta -\eta ^{n})%
\right] d\eta =0\, .
\end{equation}
Now, integration over $\sigma_i$ immediately leads to the hydrodynamic
conservation law:
\begin{equation}
f_{t}^{i}+(s(\eta ^{i},x,t)f^{i})_{x}=0 \, ,  \label{b}
\end{equation}
which is valid in the small vicinity of $x_0$. If we assume that the above
restrictions on the behaviour of functions $\eta^i(x,t)$ hold for any $x=x_0
\in \mathbb{R}$, $t>0$, equation (\ref{b}) will be valid on the entire real
line. Setting $i=1, \dots, N$ in (\ref{int1}) we immediately obtain the
first $N$ equations in (\ref{c}).

To derive the second set of equations (\ref{c}) we multiply (\ref{d}) by $%
(\eta -\eta ^{i})$ and integrate over $\sigma_j$ to get
\begin{equation}
\begin{split}  \label{int2}
\int \limits_{\sigma_j} \left[ \sum\limits_{n=1}^{N}[f_{t}^{n}+(s(\eta
,x,t)f^{n})_{x}]\delta (\eta -\eta ^{n})(\eta -\eta ^{i})\right] d\eta \\
-\int \limits_{\sigma_j} \left[ \sum\limits_{n=1}^{N}[f^{n}\eta
_{t}^{n}+s(\eta ,x,t)f^{n}\eta _{x}^{n}](\eta -\eta ^{i})\delta ^{\prime
}(\eta -\eta ^{n})\right] d\eta =0.
\end{split}%
\end{equation}
If $j=i$, the first integral in (\ref{int2}) vanishes, while the second one,
after integrating by parts and utilising the fact that each interval $%
\sigma_i$ contains only its ``own" value $\eta^i$, yields
\begin{equation}
\begin{split}  \label{int3}
\int \limits_{\sigma_i} \frac{\partial }{\partial \eta }\left( (\eta -\eta
^{i})[f^{i}\eta _{t}^{i}+s(\eta ,x,t)f^{i}\eta _{x}^{i}]\right) \delta (\eta
-\eta ^{i}) d\eta \\
=\int \limits_{\sigma_i} \left( [f^{i}\eta _{t}^{i}+s(\eta ,x,t)f^{i}\eta
_{x}^{i}]+(\eta -\eta ^{i})\frac{\partial s(\eta ,x,t)}{\partial \eta }%
f^{i}\eta _{x}^{i}\right) \delta (\eta -\eta ^{i}) d\eta =0.
\end{split}%
\end{equation}
Evaluating the integral in (\ref{int3}) we get
\begin{equation}
\eta _{t}^{i}+s(\eta ^{i},x,t)\eta _{x}^{i}=0, \quad i=1, \dots, N\, .
\label{h}
\end{equation}
It is not difficult to see that, if $j \ne i$, we recover equations (\ref{b}%
). Thus, the compatibility of the non-isospectral ansatz (\ref{05}) with the
kinetic equation (\ref{a}), (\ref{kin1}) imposes restrictions (\ref{h}) on
the functions $\eta ^{i}(x,t)$.

Generally, $2N$-component hydrodynamic type system (\ref{c}) possesses $N$
conservation laws
\begin{equation*}
\partial _{t}(\varphi _{i}(\eta ^{i})f^{i})+(s(\eta ^{i},x,t)\varphi
_{i}(\eta ^{i})f^{i})_{x}=0 ,
\end{equation*}%
where $\varphi _{i}(\eta ^{i})$ are arbitrary functions of a single
variable. It is convenient to choose $\varphi _{i}(\eta ^{i})=\eta ^{i}$ so
that (\ref{c}) reduces to (cf. (\ref{01}))%
\begin{equation}
u_{t}^{i}=(u^{i}v^{i})_{x},\text{ \ \ }\eta _{t}^{i}=v^{i}\eta
_{x}^{i},\qquad i=1,\dots ,N\,,  \label{06}
\end{equation}%
where $u^i=\eta^i f^i$, $v^{i}=-s(\eta ^{i},x,t)$.

\subsection{Closure relations}

The closure relations connecting the field variables $u^{i}$, $v^{i}$, and $%
\eta^i$ in (\ref{06}) are obtained by substituting the same ansatz (\ref{05}%
) into the integral equation (\ref{kin1}). Since we are going to use the
variables $u^i$ instead of $f^i$, we slightly modify ansatz (\ref{05}) as
follows
\begin{equation}
\eta f(\eta ,x,t)=\sum\limits_{i=1}^{N}u^{i}(x,t)\delta (\eta -\eta ^{i}) \,
.  \label{ansatz}
\end{equation}
Substitution of (\ref{ansatz}) into (\ref{kin1}) yields
\begin{equation}
s(\eta, x,t )=S(\eta )+\sum\limits_{m=1}^{N}u^{m}\frac{G(\eta ,\eta ^{m})}{%
\eta \eta ^{m}}[s(\eta ^{m}, x,t)-s(\eta, x,t )]\,,  \label{e}
\end{equation}%
As in \cite{kinetic}, we introduce (see (\ref{03}))
\begin{equation}  \label{closure1}
\epsilon ^{ik}=\frac{G(\eta ^{i},\eta ^{k}) }{\eta ^{i}\eta ^{k}}, \qquad i
\ne k \, .
\end{equation}
There is an important point to be made. In the linearly degenerate
reductions (\ref{01}), (\ref{02}) associated with the isospectral ansatz (%
\ref{cold}) involving arbitrary \textit{constants} $\eta^i$, the
dependencies of $\xi^i = - S(\eta^i)$ and $\epsilon^{ik}$ on the relevant
components of the vector ${\boldsymbol{\eta}}= \{\eta^1, \eta^2, \dots,
\eta^N \}$ are not important from the viewpoint of integrability --- these
only provide the connection with the original nonlocal equation (\ref{kin1})
--- see \cite{kinetic}. However, under the generalised ansatz (\ref{05}), $%
\eta^i$s become dependent \textit{variables}, $\eta^i = \eta^i(x,t)$, so the
aforementioned dependencies become essential for the structure of the
corresponding hydrodynamic reductions.

Now we pass to the limit as $\eta \to \eta^i$ in (\ref{e}) to obtain,
assuming $\lim _{\eta \to \eta^i } s(\eta, x, t) = - v^i$ (continuity),
\begin{equation}
v^i=\underset{m\neq i}{\sum }\epsilon ^{im}u^{m}(v^i-v^m) - S(\eta^i) +
\frac{u^{i}}{(\eta ^{i})^{2}}\underset{\eta \rightarrow \eta ^{i}}{\lim }%
G(\eta ,\eta ^{i})(s(\eta, x,t ) - s(\eta^i, x,t )).  \label{j}
\end{equation}
If the limit
\begin{equation}
\underset{\eta \rightarrow \eta ^{i}}{\lim }G(\eta ,\eta ^{i})(s(\eta, x,
t)-s(\eta^i, x, t ))  \label{limit}
\end{equation}%
exists then (\ref{j}) becomes%
\begin{equation}
v^i=\underset{m\neq i}{\sum }\epsilon ^{im}u^{m}(v^i - v^m)- S(\eta^i)+g_i(%
\mathbf{u}, \mathbf{v}, {\boldsymbol{\eta}} ),  \label{lin}
\end{equation}%
where%
\begin{equation*}
g_i(\mathbf{u}, \mathbf{v}, {\boldsymbol{\eta}} )=\frac{u^{i}}{(\eta
^{i})^{2}}\underset{\eta \rightarrow \eta ^{i}}{\lim }G(\eta ,\eta
^{i})[s(\eta, x,t)-s(\eta^i, x,t )].
\end{equation*}%
The existence of the limit (\ref{limit}) implies that the function $G(\eta
,\mu )$ has \textit{at most} a simple pole singularity on the diagonal $%
\mu=\eta$.

If the limit (\ref{limit}) vanishes for all $i=1, \dots, N$ (which happens
if $G(\eta, \mu)$ either vanishes itself or has a singularity weaker than a
simple pole as $\mu \to \eta$) then $g_i\equiv 0$ and equation (\ref{lin})
reduces to the closure conditions (\ref{02}), (\ref{03}) obtained for the
isospectral cold-gas reduction. Below we restrict our consideration just to
this, most important, case, which arises, in particular, in the case of the
kinetic equation for the KdV solitons \cite{el03}, when the kernel function $%
G(\eta, \mu)$ has only logarithmic singularity on the diagonal -- see (\ref%
{kdv}).

In conclusion of this section we note that nonexistence of the limit (\ref%
{limit}) for some given $G(\mu, \eta)$ signifies incompatibility of the
delta function ansatz (\ref{ansatz}) with the integral equation (\ref{kin1})
for that particular kernel $G(\mu, \eta)$.

\section{The structure of generalised multi-flow hydrodynamic reductions}

Motivated by the results of \cite{kinetic} for the isospectral cold-gas
hydrodynamic reductions (\ref{01}), (\ref{02}) we introduce a symmetric
matrix $\hat{\epsilon} = [\epsilon^{mn}]_{N \times N}$ with the off-diagonal
elements $\epsilon^{ik}({\boldsymbol{\eta}})$ defined by (\ref{closure1})
and the diagonal elements $\epsilon^{kk}$ being some new field variables $%
r^{k}(\mathbf{u}, {\boldsymbol{\eta}} )$.

\textbf{Theorem 1.} (\cite{kinetic}): \textit{Algebraic system} (\ref{02})
\textit{admits the parametric solution}%
\begin{equation}
u^{i}=\overset{N}{\underset{m=1}{\sum }}\beta _{mi},\text{ \ \ }v^{i}=\frac{1%
}{u^{i}}\overset{N}{\underset{m=1}{\sum }}\xi _{m}\beta _{mi},  \label{beta}
\end{equation}%
\textit{where symmetric functions} $\beta _{ik}(\mathbf{r}, {\boldsymbol{%
\eta }})$ \textit{are the elements of the matrix} $\mathbf{\hat{\beta}}%
=[\beta_{mn}]_{N \times N}$ \textit{such that} $\mathbf{\hat {\beta} \hat {%
\epsilon}=-1}$.

\textbf{Proof}: We replace (\ref{02}) by an equivalent system
\begin{equation}
v^{i}= \xi_i + \overset{N}{\underset{m=1}{\sum }}{\epsilon}
^{im}u^{m}(v^{m}-v^{i}).  \label{first}
\end{equation}%
(note that summation in (\ref{first}) goes over all $m$ including $m=i$ (cf.
(\ref{02})). Then (\ref{first}) can be re-written in the form
\begin{equation*}
v^{i}\left( 1+\overset{N}{\underset{m=1}{\sum }}\epsilon ^{im}u^{m}\right) =
\xi_i + \overset{N}{\underset{m=1}{\sum }}\epsilon ^{im}u^{m}v^{m}.
\end{equation*}%
Substituting (\ref{beta}) into the above formula we obtain
\begin{equation}  \label{param}
v^{i}\left( 1+\overset{N}{\underset{m=1}{\sum }}\overset{N}{\underset{k=1}{%
\sum }}\beta _{mk}\epsilon ^{ki}\right) = \xi_i + \overset{N}{\underset{m=1}{%
\sum }}\overset{N}{\underset{k=1}{\sum }}\xi _{m}\beta _{mk}\epsilon ^{ki}.
\end{equation}%
Taking into account $\mathbf{\hat \beta\hat{\epsilon}=-1}$, one can see that
expressions at both sides of (\ref{param}) vanish independently. Thus (\ref%
{param}) is an identity, hence the parametric representation (\ref{beta}) is
consistent with system (\ref{02}). The Theorem is proved.

\medskip \textbf{Corollary}: The field variables $r^{k}(\mathbf{u},{%
\boldsymbol{\eta }})$ are rational functions of the conserved densities $%
u^{m}$, namely,
\begin{equation}
r^{k}=-\frac{1}{u^{k}}\left( 1+\underset{m\neq k}{\sum }u^{m}\epsilon ^{mk}(%
\mathbf{\boldsymbol{\eta }})\right) ,\text{ \ }k=1,2,...,N.  \label{rim}
\end{equation}%
Indeed, multiplying both sides of the first relationship in\ (\ref{beta}) by
$\epsilon ^{ik}$ and performing summation over $i$ we obtain:
\begin{equation*}
\overset{N}{\underset{m=1}{\sum }}u^{m}\epsilon ^{mk}=\overset{N}{\underset{%
m=1}{\sum }}\overset{N}{\underset{n=1}{\sum }}\beta _{mn}\epsilon ^{nk}=-1.
\end{equation*}%
Thus,%
\begin{equation*}
\underset{m\neq k}{\sum }u^{m}\epsilon ^{mk}+r^{k}u^{k}=-1,
\end{equation*}%
which immediately yields (\ref{rim}).

\medskip

\textbf{Theorem 2}: \textit{Under parametrization }(\ref{beta}) \textit{the $2N$-component
hydrodynamic type system} (\ref{06}), (\ref{02}), (\ref{03}) \textit{reduces
to a quasi-diaginal form:}%
\begin{eqnarray}
\eta _{t}^{i} &=&v^{i}\eta _{x}^{i}, \quad i=1, \dots, N;  \label{r1} \\
r_{t}^{k}&=&v^{k}r_{x}^{k}+\frac{1}{u^{k}}\left( \underset{n\neq k}{\sum }
u^{n}(v^{n}-v^{k})\frac{\partial \epsilon ^{nk}}{\partial \eta ^{k}}-\xi
_{k}^{\prime }\right) \eta _{x}^{k}, \quad k=1, \dots, N \, . \label{r2}
\end{eqnarray}

\textbf{Proof}: The evolution equations (\ref{r1}) for $\eta^i(x,t)$ are the
same as in (\ref{06}) so we need only to derive equations (\ref{r2}) for $%
r^k(x,t)$, $k=1, \dots, N$. Substituting parametric representation (\ref%
{beta}) into the conservation laws (\ref{06}) we obtain
\begin{equation*}
\partial _{t}\left( \overset{N}{\underset{m=1}{\sum }}\beta _{mi}\right)
=\partial _{x}\left( \overset{N}{\underset{m=1}{\sum }}\xi _{m}\beta
_{mi}\right) .
\end{equation*}
Multiplying both sides by $\epsilon ^{ik}$, performing summation over the
repeated index $i$ and using the relationship $\mathbf{\hat{\beta}\hat{%
\epsilon}=-1}$, one arrives at the equation
\begin{equation}  \label{equ}
\overset{N}{\underset{i=1}{\sum }}\overset{N}{\underset{m=1}{\sum }}\beta
_{mi}\partial _{t}\epsilon ^{ik}=\overset{N}{\underset{i=1}{\sum }}\overset{N%
}{\underset{m=1}{\sum }}\xi _{m}\beta _{mi}\partial _{x}\epsilon
^{ik}-\partial _{x}\xi _{k}.
\end{equation}%
A simple but not entirely trivial calculation using the following obvious
property of the matrix $\hat \epsilon (\mathbf{r}, {\boldsymbol{\eta}})$:
\begin{equation*}
\frac{\partial \epsilon^{nk}}{\partial r^s} = \delta_{nk}\delta_{ks}\, ,
\end{equation*}
and the evolution equations (\ref{r1}) for $\eta^k$, shows that (\ref{equ})
reduces to
\begin{equation}\label{r20}
r_{t}^{k}=v^{k}r_{x}^{k}+ \frac{1}{u^{k}}\overset{N}{\underset{s=1}{\sum }}%
\left( \overset{N}{\underset{n=1}{\sum }}\overset{N}{\underset{m=1}{\sum }}%
\beta _{mn}(\xi _{m}-v^{s}) \frac{\partial \epsilon ^{nk}}{\partial \eta ^{s}%
}-\frac{\partial \xi _{k}}{\partial \eta ^{s}}\right) \eta _{x}^{s}, \ \
k=1, \dots, N.
\end{equation}
Then taking into account (see (\ref{beta}))
\begin{equation*}
\overset{N}{\underset{m=1}{\sum }}\beta _{mi}=u^{i},\text{ \ \ }\overset{N}{%
\underset{m=1}{\sum }}\xi _{m}\beta _{mi}=u^{i}v^{i},
\end{equation*}%
equations (\ref{r20}) reduce to the form (\ref{r2}).
The Theorem is proved.

\medskip

Eliminating $u^i$ from (\ref{beta}), we obtain expressions relating $\mathbf{%
v}$ to ${\boldsymbol{\eta}}$ and $\mathbf{r}$:
\begin{equation}  \label{vi}
v^i(\mathbf{r}, {\boldsymbol{\eta}})=\frac{\overset{N}{\underset{m=1}{\sum }}%
\xi_{m}\beta_{mi}}{\overset{N}{\underset{m=1}{\sum }}\beta_{mi}}\, ,
\end{equation}
while $u^i(\mathbf{r}, {\boldsymbol{\eta}})$ are given by the first equation
in (\ref{beta}). Now, system (\ref{r1}), (\ref{r2}) is closed.

For the isospectral case, when $\eta^i$, $i=1, 2, \dots, N$ are constants so
$\partial_t \eta^i = \partial_x \eta^i =0$ and we recover from (\ref{r2})
the Riemann invariant representation
\begin{equation}
r_{t}^{i}=v^{i}(\mathbf{r})r_{x}^{i}  \label{diag}
\end{equation}
of system (\ref{01}), (\ref{02}) obtained in \cite{kinetic} with the use of
the linear degeneracy of the isospectral cold-gas hydrodynamic reductions.
As a matter of fact, one can see now that the Riemann invariant equations (%
\ref{diag}) could be readily obtained directly from system (\ref{01}), (\ref%
{02}) by the substitution into it of the parametric solution (\ref{beta}).

Thus, the $2N$-component hydrodynamic reduction (\ref{06}) admits
parametrization (\ref{beta}) resolving algebraic system (\ref{02}) and
reducing the evolution equations to the form (\ref{r1}), (\ref{r2}). System (%
\ref{r1}), (\ref{r2}) has double characteristic velocities $v^{k}(\mathbf{\ r%
}, {\boldsymbol{\eta }})$ (\ref{vi}). However, in the general case, {\bf just $N$
functions} $\eta ^{k}(x,t)$ are Riemann invariants (i.e. only a half of the
complete hydrodynamic system (\ref{06}) can be written in diagonal form),
while the field variables $r^{k}(x,t)$ become Riemann invariants only if the {\bf corresponding}
$\eta ^{k}=$const.

In conclusion of this Section we note that linear degeneracy of system (\ref%
{diag}) proved in \cite{kinetic} implies that $\partial v^i(\mathbf{r}%
)/\partial r^i=0$ for all $i=1, \dots N$. The latter property clearly
remains valid for the characteristic velocities $v^i( \mathbf{r}, {%
\boldsymbol{\eta}})$ of the generalised hydrodynamic reductions (\ref{r1}), (%
\ref{r2}), however, now this is no longer associated with the notion of
linear degeneracy of a hydrodynamic type system in the classical sense \cite%
{lax}, \cite{yan} since $r^k$ are no longer Riemann invariants and also $%
\partial v^i(\mathbf{r}, {\boldsymbol{\eta}})/\partial \eta^i \ne 0$.

\section{Conclusion}

In this paper, we have derived the generalised hydrodynamic reductions of
the nonlocal kinetic equation for a soliton gas (\ref{a}), (\ref{kin1}) by
considering the non-isospectral multi-flow ansatz (\ref{05}) for the
distribution function. These new reductions have turned out to have rather
unusual structure which we have revealed by using the parametric solution (%
\ref{beta}) to the algebraic closure conditions (\ref{02}), (\ref{03}). More
precisely, the non-isospectral $N$-flow hydrodynamic reductions of the
kinetic equation are shown to represent $2N$-component half-diagonal systems
of hydrodynamic type (\ref{r1}), (\ref{r2}) with $N$ Riemann invariants and $%
N$ double characteristic velocities. The feature that makes the derived
reductions deserving special attention is that, while they are clearly not
integrable by Tsarev's generalised hodograph transform method \cite{ts91},
they could still prove to be integrable in some new sense yet to be
understood. Indeed, having in mind that system (\ref{r1}), (\ref{r2}) can be
derived as a generalised hydrodynamic reduction of the kinetic equation
associated with an integrable equation (e.g. with the KdV equation --- for $%
S(\eta)$, $G(\eta, \mu)$ defined by (\ref{kdv})),  one can expect that this
reduction will be integrable by some nontrivial extension of the generalised
hodograph method.

As a by-product of our calculations we recover the Riemann invariant
structure of the isospectral cold-gas reductions (\ref{01}), (\ref{02}), (%
\ref{03}) studied in \cite{kinetic}. We note that our present compact
derivation, unlike that in \cite{kinetic}, does not make any use of the
linear degeneracy property of the reductions under study.

\section*{Acknowledgement}

We thank Alexander Chesnokov for stimulating and clarifying discussions. We also indebted to Mikhail Zhukov
for pointing out the possibility of reducing
equation (\ref{r20}) to the quasi-diagonal form (\ref{r2}).

MVP's work was partially supported by the RF Government grant
\#2010-220-01-077, ag. \#11.G34.31.0005, by the grant of Presidium of RAS
``Fundamental Problems of Nonlinear Dynamics''\ and by the RFBR grant
11-01-00197. MVP is also grateful to the SISSA, Trieste (Italy) where some
part of this work has been done.

\end{document}